\documentclass[aps,onecolumn,pra,preprintnumbers,superscriptaddress,amsmath,amssymb]{revtex4-1}
\usepackage{graphicx}
\usepackage{dcolumn}
\usepackage{bm}
\usepackage{tikz} \usepackage{mathtools}
\newcommand*\diff{\mathop{}\!\mathrm{d}} 
\usepackage{braket} \usepackage[caption=false]{subfig} \usepackage{natbib}
 \usepackage{hyperref}
\hypersetup{ colorlinks, allcolors=[rgb]{.18,.19,.57} }
 
\usepackage[textsize=tiny]{todonotes} \usepackage{setspace} \makeatletter
\renewcommand{\todo}[2][]{%
  \@todo[caption={#2}, #1]{\begin{spacing}{0.5}#2\end{spacing}}%
}\makeatother 

\begin{document}
\title{Optimized Raman pulses for atom
  interferometry}
\author{Jack Saywell} \email[]{j.c.saywell@soton.ac.uk} \author{Max Carey}
\email[]{max.carey@soton.ac.uk} \author{Mohammad Belal} \affiliation{School of
  Physics \& Astronomy, University of Southampton, Highfield, Southampton, SO17
  1BJ, UK} \author{Ilya Kuprov} \affiliation{School of Chemistry, University of
  Southampton, Highfield, Southampton, SO17 1BJ, UK} \author{Tim Freegarde}
\affiliation{School of Physics \& Astronomy, University of Southampton,
  Highfield, Southampton, SO17 1BJ, UK} \date{\today}

\begin{abstract}
  We present mirror and beamsplitter pulse designs that improve the fidelity of
  atom interferometry and increase its tolerance of systematic inhomogeneities.
  These designs are demonstrated experimentally with a cold thermal sample of
  $^{85}$Rb atoms. We first show a stimulated Raman inversion pulse design that
  achieves a ground hyperfine state transfer efficiency of 99.8(3)\%, compared
  with a conventional $\pi$ pulse efficiency of 75(3)\%. This inversion pulse is
  robust to variations in laser intensity and detuning, maintaining a transfer
  efficiency of 90\% at detunings for which the $\pi$ pulse fidelity is below
  20\%, and is thus suitable for large momentum transfer interferometers using
  thermal atoms or operating in non-ideal environments. We then extend our
  optimization to all components of a Mach-Zehnder atom interferometer sequence and
  show that with a highly inhomogeneous atomic sample the fringe visibility is increased
  threefold over that using conventional $\pi$ and $\pi/2$ pulses.
\end{abstract}

\maketitle

\section{\label{sec:Introduction}Introduction}

Atom interferometers \cite{RBerman1997} are the matterwave analogues of optical
interferometers. Slow, massive atomic wavepackets replace the photons that are
divided to follow separate spatial paths before being recombined to produce
interference; and, in place of the mirrors and beamsplitters, carefully-timed
resonant laser pulses split, steer and recombine the wavepackets. Atom
interferometers have already demonstrated unprecedented performance for inertial
measurement, with potential applications from navigation \cite{Canuel2006,
  Dickerson2013,Barrett2014, Cheiney2018} to the detection of gravitational
waves \cite{Dimopoulos2009, Graham2013} and investigations of dark energy
\cite{Hamilton2015, Burrage2015}.

As with an optical interferometer, the sensitivity of an atom interferometer
depends upon the lengths and separation of the interfering paths and the
coherence and number of quanta detected. Whereas optical interferometers are
possible on the kilometre scale using ultra-stable lasers and optical fibre
components, the path separations in atom interferometers result from momentum
differences of only one or a few photon recoils, and expansion of the atom cloud
limits the interferometer duration. Large momentum transfer (LMT)
interferometers increase the path separation by employing repeated augmentation
pulses to impart multiple photon impulses \cite{McGuirk2000}, but any
sensitivity improvements thus achieved risk an accompanying reduction in fringe
visibility as losses quickly accrue when the optical interactions are imperfect \cite{Butts2013, Szigeti2012}. LMT
interferometers typically rely on an atomic sample with a narrow initial
momentum distribution \cite{Szigeti2012,kovachy2015quantum}, with Bloch
oscillations \cite{McDonald2013} and Bragg diffraction \cite{Muller2008,
  Altin2013, Kovachy2015} demonstrating the greatest separation, but filtering the atomic
sample in this way to reduce the effects of inhomogeneities and cloud expansion
involves lengthier preparation and causes a fall in the signal-to-noise ratio
because fewer atoms are measured.

For applications such as inertial navigation where both the sensitivity and
repetition rate are important, techniques are required that are more tolerant of
experimental and environmental inhomogeneities in laser intensity, magnetic
field, atom velocity and radiative coupling strength. Adiabatic transfer
\cite{Melinger1994, Garwood2001, Kovachy2012a, Kotru2015, Jaffe2018} allows robust, high-fidelity state transfer, but is
necessarily a slow process not suited to preparing or resolving superpositions
\cite{Bateman2007}. Composite and shaped pulses \cite{Levitt1981a, Levitt1986, Freeman1998a, McDonald1991} are attractive
alternatives. Originally developed for nuclear magnetic resonance (NMR)
spectroscopy, composite pulses are concatenated sequences of pulses with
tailored phases and durations that can replace the fractional Rabi oscillations in
simple atom interferometers and increase the tolerance of inhomogeneities in the atom-laser interaction \cite{Dunning2014, Butts2013}.

These pulse shapes can be tailored and improved using optimal
control techniques \cite{Warren1993, Brif2010, Glaser2015} that have been applied
extensively in NMR \cite{Skinner2003a, Khaneja2005, Kobzar2004, Hogben2011, DeFouquieres2011, Kobzar2012, Goodwin2015}, and have also been successfully demonstrated in nitrogen-vacancy (NV) magnetometry
\cite{Nobauer2015}, ultra-cold molecule stabilization \cite{Koch2004} and the
control of Bose-Einstein condensates \cite{VanFrank2014, Jager2014, Wigley2016}.

We have previously investigated the application of optimal control techniques to the
optimization of mirror pulses for interferometry, showing computationally how
this can maximize interferometer contrast by compensating for realistic
experimental inhomogeneities in detuning and coupling strength
\cite{Saywell2018b}. We now build on this, presenting the design and
experimental implementation of a high-fidelity inversion pulse and a novel
approach to optimizing an entire 3-pulse interferometer sequence.

Our inversion pulse achieves 99.8(3)\% transfer between the two hyperfine ground
states in a thermal sample of $^{85}$Rb where a rectangular $\pi$ pulse achieves
only 75(3)\%, with a greater velocity acceptance than existing composite and
shaped pulses making it particularly suited for LMT applications. Our optimized
3-pulse interferometer demonstrates a threefold increase in the experimental
fringe visibility with a 94(4) $\mu$K atom sample compared to a conventional
Mach-Zehnder interferometer using rectangular pulses. This is, to our knowledge, the first demonstration
of shaped individual beamsplitter pulses preparing momentum superpositions being used to
improve the contrast of an atom interferometer.

\section{Theoretical system and optimization approach} \label{sec:optimal-pulse-design}

We consider an alkali atom undergoing stimulated Raman transitions between
hyperfine levels, forming an effective 2-level system described using the basis
states $\ket{\mathrm{g},\bm{\mathrm{p}}}$, and $\ket{\mathrm{e},
  \bm{\mathrm{p}}+\hbar \bm{\mathrm{k}}_L}$ \cite{Kasevich1991a}.
$\bm{\mathrm{p}}$ is the atomic momentum and $\bm{\mathrm{k}}_L$ represents the
difference between the wave vectors of the two lasers, or `effective' wave
vector. The change in state under the action of a pulse of constant intensity
and combined laser phase $\phi_L$, acting for duration $\Delta$t, is described
by a propagator
\begin{equation} \label{eq: prop} \hat{U} = \begin{pmatrix}
    C^* & -iS^* \\
    -iS & C
  \end{pmatrix},
\end{equation}
where $C$ and $S$ are defined as \cite{Stoner2011}:
\begin{flalign} \label{eq: propagator_elements} \nonumber
  C &\equiv \cos\bigg(\frac{\sqrt{\Omega_R^2 + \delta^2}\Delta t }{2}\bigg) \\
  &\quad+ i\bigg(\frac{\delta} { \sqrt{\Omega_R^2 + \delta^2}}\bigg)\sin\bigg(\frac{\sqrt{\Omega_R^2 + \delta^2}\Delta t} {2}\bigg)\\
  S &\equiv e^{i\phi_L}\bigg(\frac{{\Omega}_R } { \sqrt{\Omega_R^2 +
      \delta^2}}\bigg)\sin\bigg(\frac{\sqrt{\Omega_R^2 + \delta^2}\Delta t
  }{2}\bigg).
\end{flalign}
$\Omega_R$ is the two-photon Rabi frequency, and $\delta$ is the two-photon
Raman detuning \cite{RBerman1997}, which depends on atomic momentum and is
assumed to be approximately constant for the duration of the pulses. This is
often achieved by chirping the frequency difference of the Raman beams to
account for the Doppler shift caused by gravitational acceleration
\cite{Kasevich1991a}.

Rectangular $\pi/2$ and $\pi$ pulses are the building blocks of conventional
interferometer sequences, and result from fixing the duration of a pulse with
constant laser intensity and frequency such that the quantum state undergoes a $\pi/2$ or
$\pi$ rotation about an axis in the xy plane of the Bloch sphere
\cite{Feynman1957}. Variations in the detuning (e.g. due to thermal motion)
and Rabi frequency (e.g. due to beam intensity variation) across the atom cloud
degrade the interferometer signal \cite{Gauguet2009, Hoth2016}. In the NMR literature
these errors are referred to as `off-resonance' and `pulse-length' errors
respectively. If the individual pulses no longer perform the intended operations
for each atom, the contrast of the interferometer fringes falls, their offset
varies, and the inertial phase is modified \cite{Stoner2011, Luo2016,
  Saywell2018b}. Although the problem of intensity inhomogeneity in atom
interferometry may be compensated by the use of collimated top-hat laser beams
\cite{Mielec2018}, our approach obtains tailored pulses that achieve mutual
compensation of inhomogeneous coupling strengths and large Raman detunings
without the need for additional optical elements.

We define our pulses in terms of multiple discrete time `slices' where the
combined Raman laser phase $\phi_L$ takes a different value for each equal
timestep $\diff t$ in the pulse. They are therefore described by profiles $\phi_L(t)
= \{\phi_0,\phi_1,...,\phi_N\}$, with a total duration $\tau_{\mathrm{pulse}}$.
Although we could also choose to vary the pulse amplitude with time, for experimental simplicity we have
considered pulses with constant amplitude profiles in this work. The laser phase
forms our control parameter, and the action of a given pulse on our effective
two-level system may be described by a sequence of pulse propagators
$\hat{U}_n(\phi_n, \Omega_R, \delta, \diff t)$. The action of a pulse is then a
time-ordered product of propagators
\begin{equation}
  \hat{U} = \hat{U}_N \hat{U}_{N-1} \ldots \hat{U}_{n} \ldots \hat{U}_{1}   \hat{U}_{0},
\end{equation}
where the propagator for the nth timestep $\hat{U}_n$ takes the form of a
rectangular pulse of fixed laser amplitude and phase acting for duration $\diff t$
(Equation \ref{eq: prop}) \cite{Stoner2011, Saywell2018b}.

Optimal control theory finds the `best' way to control the evolution of a
system so as to maximize some fidelity or `measure of performance'. Often, this
fidelity is taken to be the accuracy with which initial states are driven to
target states by the optimal modulation of available control fields. We employ
the gradient ascent pulse engineering (GRAPE) \cite{Khaneja2005} algorithm to
design optimal pulses for our purposes. Given an initial guess for the pulse and
a choice of pulse fidelity, GRAPE efficiently calculates the required propagator
derivatives and recent improvements permit the use of a fast second order
optimization, known as the limited-memory Broyden-Fletcher-Goldfarb-Shanno
(L-BFGS) quasi-Newton method \cite{DeFouquieres2011}. By defining an ensemble of
systems with a distribution of atomic detunings (giving rise to `off-resonance'
errors) and variations in coupling strength (or, equivalently, `pulse-length'
errors), a robust pulse that maximizes the chosen fidelity over the ensemble may
be obtained by averaging over the ensemble in the fidelity calculation. The length and number of timesteps are chosen at the outset and fixed when optimizing a pulse. Longer pulses can typically achieve higher terminal fidelities \cite{Kobzar2004, Kobzar2012}. The spin
dynamics simulation software Spinach \cite{Hogben2011}, and its optimal control
module, was used to optimize pulses in this work.

\section{Measures of pulse performance}\label{sec:fidelities}

The choice of pulse fidelity used in the optimization depends on the
application, and requires a careful consideration of the experimental
requirements. For example, we may write our pulse fidelity, $\mathcal{F}$, for a
given atom as a function of the overlap of a chosen target state $\ket{\psi_T}$
with the final state after application of the pulse to our initial state
$\ket{\psi_0}$: $\mathcal{F} = |\bra{\psi_T}\hat{U}\ket{\psi_0}|^2$. This
fidelity, when maximized, gives us a state-transfer or `point-to-point' (PP)
pulse \cite{Kobzar2004}. Alternatively, the goal of the optimization may be to
recreate a specific target propagator $\hat{U}_T$ and we consider the fidelity
to be $\mathcal{F} = \frac{1}{2}\mathrm{Tr}(\hat{U}_T^{\dagger} \hat{U})$, yielding a
so-called `universal rotation' (UR) pulse \cite{Kobzar2012}. Many other choices of pulse fidelity are available, however, including those which map a range of initial states to different targets, and are not aimed at obtaining full universal rotations \cite{VanFrank2014}. We discuss appropriate fidelity choices that maximize fringe visibility and minimize
any unwanted spread in the inertial phase for two cases. Firstly, we consider a
fidelity choice for pulses used to impart additional momentum in extended LMT
interferometer sequences. Secondly, we present measures of performance for each pulse within
a three-pulse `Mach-Zehnder' interferometer sequence.

In LMT interferometers, the beamsplitter and mirror operations are extended by
multiple `augmentation' pulses with alternating effective wavevectors designed
to swap the population of the internal states whilst imparting additional
momentum \cite{McGuirk2000, Butts2013, Kotru2015}. In order to optimize an augmentation pulse
for LMT interferometry, which we represent by the propagator $\hat{U}_{\mathrm{A}}$, it may be sufficient to consider the `point-to-point'
fidelity
\begin{equation}
  \mathcal{F}_{\mathrm{A}} = |\bra{\mathrm{e}}\hat{U}_{\mathrm{A}}\ket{\mathrm{g}}|^2
\end{equation}
without concern for the relative phase introduced between the two states. This
is because, in LMT interferometers, the augmentation pulses appear in pairs
within the extended pulse sequence \cite{McGuirk2000, Butts2013, Dunning2014} so
that the interferometer phase introduced by each pulse is, to first order,
cancelled out by that introduced by a subsequent one. Choosing to optimize a PP operation as opposed to a UR pulse effectively gives the GRAPE algorithm a larger target to shoot at, allowing impressive fidelity to be achieved with a modest pulse area.

In a three-pulse Mach-Zehnder interferometer sequence, $\pi/2$ and $\pi$ pulses are applied
to atoms initially in the state $\ket{\mathrm{g}}$ separated by equal `dwell
times' for which the fields are extinguished. Inertial effects such as rotations
and accelerations imprint a relative phase $\Phi$ between the internal states at
the end of the interferometer, which is mapped onto a population difference by
the final $\pi/2$ pulse. The resulting excited state probability is
\begin{equation}
  P_{\mathrm{e}} = \frac{1}{2}(\mathcal{A}-\mathcal{B}\cos(\Phi + \Delta\phi))
\end{equation}
where $\mathcal{A}$ and $\mathcal{B}$ are the offset and contrast of the
interferometer fringes respectively. $\Delta\phi$ represents a fixed shift to
the inertial phase used to scan the fringe pattern, and must be constant from
atom to atom.

An optimal 3-pulse interferometer sequence, represented by propagators
$\hat{U}_1,\hat{U}_2,\hat{U}_3$ should maximize the contrast $\mathcal{B}$,
minimize any unwanted variation of/in the inertial phase, $\Delta\phi$, and fix
the offset $\mathcal{A}$ for all atoms with the range of detunings and coupling
strengths found in the atomic cloud. The optimal contrast and offset are
achieved for all atoms if the following conditions on the 3 pulses are met:
\begin{subequations}
  \begin{align}
    |\braket{\mathrm{e}|\hat{U}_{i =1,3}|\mathrm{g}}|^2 &= \frac{1}{2},\\
    |\braket{\mathrm{e}|\hat{U}_2|\mathrm{g}}|^2 &= 1.
  \end{align}
\end{subequations}
We interpret these conditions as requiring our beamsplitter pulses (pulses 1 and
3) to perform a $90^{\circ}$ rotation of the quantum state about an axis in the
xy plane of the Bloch sphere. Similarly, the optimal mirror pulse (pulse 2) must
perform a robust $180^{\circ}$ rotation of the quantum state. It is crucial for
interferometry that the inertial phase $\Phi$ is not modified by a different
amount for each atom as a result of the pulse sequence. This is equivalent to
requiring the combined phase shift due to the pulse sequence, $\Delta\phi$, to
be fixed or cancelled by the pulses for every atom, where
\begin{equation}
  \begin{aligned}
\Delta\phi &= \phi(\braket{\mathrm{e}|\hat{U}_1|\mathrm{g}})  -\phi(\braket{\mathrm{g}|\hat{U}_1|\mathrm{g}}) \\
  &\quad-2\phi(\braket{\mathrm{e}|\hat{U}_2|\mathrm{g}}) \\
  &\quad+ \phi(\braket{\mathrm{g}|\hat{U}_3|\mathrm{g}}) + \phi(\braket{\mathrm{e}|\hat{U}_3|\mathrm{g}}).
  \end{aligned}
\end{equation}
We use the notation $\phi(\braket{a|b})$ to indicate the argument of the overlap
$\braket{a|b}$. If $\Delta\phi$ varies from atom to atom as it does with some
pulses \cite{Luo2016, Saywell2018b}, the resulting contrast after thermal
averaging will be washed out. A careful choice of pulse fidelity for the
beamsplitters and mirrors will lead to pulses that maximize interferometer
contrast and minimize unwanted variation in the interferometer phase $\Phi$ from
atom to atom.

Our first interferometer pulse $\hat{U}_1$ is designed to take atoms from the
initial ground state, to a balanced superposition with well-defined phase on the
equator of the Bloch sphere, $\ket{\psi_T} = (\ket{\mathrm{g}} +
\ket{\mathrm{e}})/\sqrt{2}$. This means our beamsplitter fidelity,
$\mathcal{F}_1$, can be written as
\begin{equation}
  \mathcal{F}_1 = |\bra{\psi_T}\hat{U}_1\ket{\mathrm{g}}|^2,
\end{equation}
yielding a PP 90$^{\circ}$ pulse. This choice of fidelity, if maximized, results in a beamsplitter which satisfies the conditions
$|\braket{\mathrm{e}|\hat{U}_1|\mathrm{g}}|^2 = 1/2$ and
$\phi(\braket{\mathrm{e}|\hat{U}_1|\mathrm{g}})
-\phi(\braket{\mathrm{g}|\hat{U}_1|\mathrm{g}}) = 0$.

The mirror pulse, $\hat{U}_2$, acting after the first period of free evolution,
is designed to swap the internal states, but introduce no relative phase between
them, satisfying conditions $|\braket{\mathrm{e}|\hat{U}_2|\mathrm{g}}|^2 = 1$ and
$\phi(\braket{\mathrm{e}|\hat{U}_2|\mathrm{g}}) = \mathrm{constant}$. This pulse is designed to
perform a $\pi$ rotation on the Bloch sphere about a fixed axis in the xy plane.
We can therefore consider the mirror pulse fidelity to be a measure of how close
our pulse propagator is to that for an ideal $\pi$ rotation, $\hat{U}_{\pi}$, and optimize the UR
180$^{\circ}$ fidelity
\begin{equation}
  \mathcal{F}_2 = \frac{1}{2}\mathrm{Tr}(\hat{U}^{\dagger}_{\pi}\hat{U}_2)
\end{equation}
\cite{Kobzar2012, Saywell2018b}. Designing the mirror pulse as a universal
rotation means that variations to the inertial phase $\Phi$ that vary with
$\delta$ and $\Omega_R$ are minimized. We also note that if the pulse profile of
the mirror pulse is made to be odd or antisymmetric about its temporal midpoint, then any modification to the inertial phase will be constant for all $\delta$ and
$\Omega_R$. This follows from the fact that pulses with this symmetry fix the
axis of rotation to the $xz$ plane of the Bloch sphere for all resonance
offsets, a property known in the NMR literature
\cite{Tycko1985,Odedra2012,Kobzar2012}. Antisymmetric UR 180$^{\circ}$ pulses
may be constructed by first optimizing a single PP $90^{\circ}$ pulse and
following the steps outlined by Luy \textit{et al.} \cite{Luy2005}.

Finally, the third pulse is designed to accurately map the relative phase
acquired between the two internal states at the end of the second dwell time,
onto a difference in atomic population. This pulse does not need to be a universal 90$^{\circ}$ rotation, as only the z-component of the final Bloch vector matters when measuring the excited state population at the end of the interferometer. The action of this pulse can be thought
of as a phase sensitive $\pi/2$ rotation, and may be obtained by taking the
pulse profile for the first pulse $\phi_L(t)$, and reversing it in time about
the temporal mid-point and inverting the pulse profile to obtain
$-\phi(\tau-t)$. This ``flip-reverse'' operation relies on the symmetry
properties of spin-$\frac{1}{2}$ propagators \cite{Levitt2008a, Braun2014}. The
result is a pulse which satisfies $|\braket{\mathrm{e}|\hat{U}_3|\mathrm{g}}|^2 =
1/2$ but which has $ \phi(\braket{\mathrm{g}|\hat{U}_3|\mathrm{g}}) +
\phi(\braket{\mathrm{e}|\hat{U}_3|\mathrm{g}}) =
-\phi(\braket{\mathrm{e}|\hat{U}_1|\mathrm{g}})
+\phi(\braket{\mathrm{g}|\hat{U}_1|\mathrm{g}}) + \pi$, thus satisfying the condition on
$\Delta\phi$. Optimizing the 3-pulse interferometer with these symmetry
constraints (the final pulse has the time-reversed and
inverted profile of the first beamsplitter) minimizes the unwanted modification to the inertial phase term
$\Phi$ and maximizes the contrast of resulting fringes.

\section{Results of optimizations}

\begin{figure}[t]
  \centering \includegraphics[width=246pt]{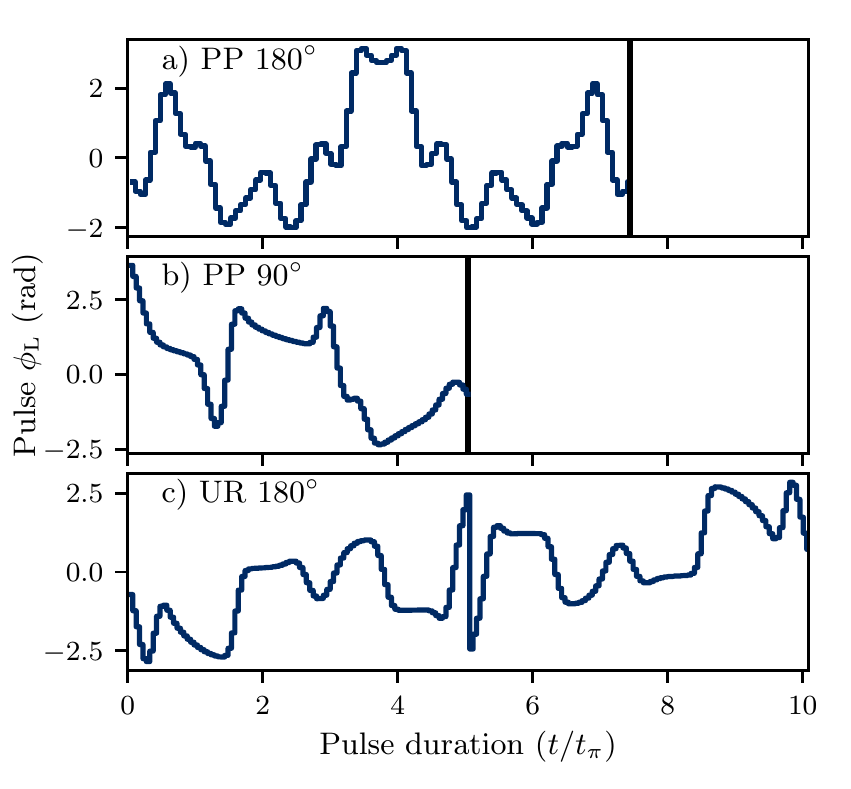}
  \caption{Optimization results for an individual LMT augmentation pulse
    $\mathcal{F}_{\mathrm{A}}$ (a), a beamsplitter pulse $\mathcal{F}_{1}$ (b),
    and an antisymmetric mirror pulse $\mathcal{F}_{2}$ (c). The pulse profiles
    are plotted against time as a fraction of the duration of a rectangular
    $\pi$ pulse, $t_{\pi}$. Each optimization was continued until fidelities
    greater than 0.99 were reached. Each pulse was optimized for an ensemble of
    atoms with a temperature of 120 $\mu$K and a range of coupling strengths of
    $\pm$ 10\% $\Omega_{\mathrm{eff}}$. }
  \label{fig:pulse_profiles}
\end{figure}

Figure \ref{fig:pulse_profiles} shows pulses obtained using GRAPE optimizing a
PP 180$^{\circ}$ inversion pulse, a PP 90$^{\circ}$ beamsplitter pulse, and an
antisymmetric UR 180$^{\circ}$ mirror pulse. The duration of each pulse, the
number of timesteps, and the initial guess for the pulse, are chosen in advance. The pulse duration and timestep number are chosen so that a sufficiently high terminal fidelity (0.99) can be achieved. The chosen fidelity is averaged over an ensemble to obtain robust solutions. The
ensembles are defined in terms of a sample of off-resonance and pulse-length
errors. The sample of off-resonance errors is taken to reflect the variation in
detuning caused by the momentum distribution of the atoms at a given temperature
along the Raman beam axis, which we assume to follow a Maxwell-Boltzmann
distribution. Choosing to optimize over an ensemble representing a larger
temperature will result in a pulse that is robust over a larger range of
detunings, and will therefore still have high fidelity if the temperature of the
atomic ensemble is reduced.

\begin{figure*}[!ht]
  \subfloat[]{\includegraphics[width=246pt]{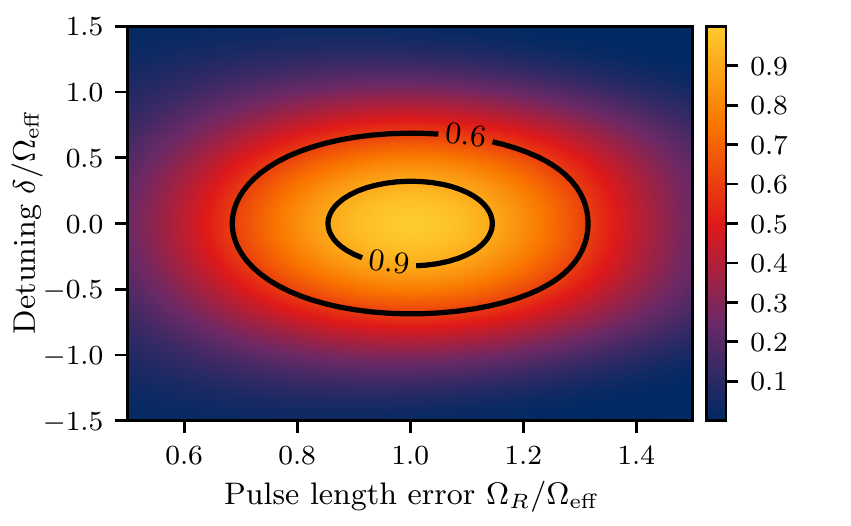}} \hfill
  \subfloat[]{\includegraphics[width=246pt]{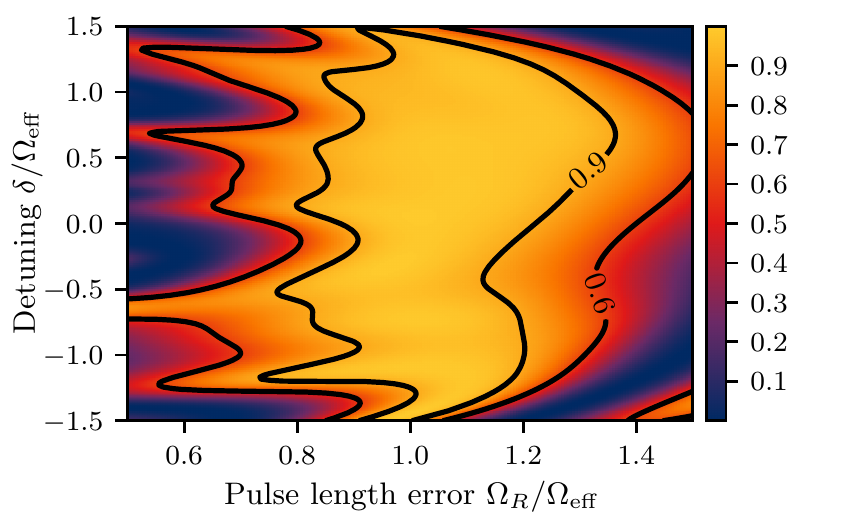}}\\
  \caption{Simulated 3-pulse interferometer contrast $\mathcal{B}$ for (a)
    standard rectangular pulses and (b) GRAPE sequence, computed for a range of
    off-resonance and pulse-length errors. Contours are at 0.6, and 0.9
    respectively. The effective Rabi frequency $\Omega_{\mathrm{eff}}$ and $\pi$
    pulse duration $t_{\pi}$ are defined by requiring that
    $\Omega_{\mathrm{eff}}t_{\pi}=\pi$. The robustness of the GRAPE sequence to
    these inhomogeneities is shown by the increased area of high contrast
    centered on resonance.}
  \label{fig:Contour}
\end{figure*}

Taking the beamsplitter and mirror pulses shown in Figure
\ref{fig:pulse_profiles}, we can simulate how the resulting interferometer
contrast varies with detuning and variation in coupling strength. Figure
\ref{fig:Contour} compares the simulated interferometer contrast $\mathcal{B}$
obtained with our ``flip-reverse'' GRAPE sequence, with the standard rectangular
$\pi/2-\pi-\pi/2$ sequence, over a range of pulse-length errors and detunings.
We find that the interferometer contrast with our optimized GRAPE pulses is more
resilient to variations in detuning and coupling strength.

\section{Experimental procedure}

We have implemented our pulses in our experimental setup, a description of
which can be found in previous work \cite{Carey2017, Dunning2014}, but we take a
moment here to outline the most salient points.

We realize our pulses on a thermal cloud of $\sim10^7$ $^{85}$Rb, released from
a 3-D magneto-optical trap (MOT), cooled in an optical molasses for $\sim 5~$ms
and optically pumped into the $5S_{1/2}, F = 2$ state with a distribution over
the five $m_F$ sublevels which, as the pumping is performed with the MOT light
along all axes, is assumed to be uniform \cite{Dunning2014}.

The cloud temperature is tuned by adjusting the intensity of the cooling light
during the optical molasses and is measured by performing Raman Doppler
spectroscopy with the Raman beams at low power \cite{Kasevich1991b}. The
temperature can be adjusted in the range of $20-200~\mu$K and, combined with the
multiple Rabi frequencies due to the different coupling strengths of the five
$m_F$ levels present in the sample, provides an inhomogeneous system with which
to explore the performance of pulses designed to provide robustness against the
resulting off-resonance and pulse-length errors.

Interferometry pulses are realised via horizontal, counter-propagating beams
with orthogonal linear polarizations that interact with the
released cloud to drive two-photon Raman transitions on the D2 line between the
$F = 2$ and $F = 3$ hyperfine ground states (states $\ket{\mathrm{g}}$ and
$\ket{\mathrm{e}}$ respectively). Both laser fields are far detuned from an
intermediate state $5P_{3/2}$, thus allowing our atoms to be modelled as
effective two-level systems \cite{RBerman1997}, and the polarization
arrangement removes the $m_F$ dependence of the light shift \cite{Dunning2014}.

One beam is formed from the first diffracted order of a 310MHz acousto-optic
modulator (AOM) and the other from the carrier suppressed output of a 2.7GHz
electro-optic modulator (EOM). The modulator frequencies sum to the hyperfine
splitting between the ground states plus a detuning $\delta$ applied to the
carrier frequency of the EOM RF signal. The beams are combined on a single AOM
to shutter the interaction light on and off with a rise-time of $\sim 100~$ns
before being separated by polarization and delivered to the atom cloud through
separate polarization-maintaining fibers.

\begin{figure}[b]
  \centering \includegraphics[width=246pt]{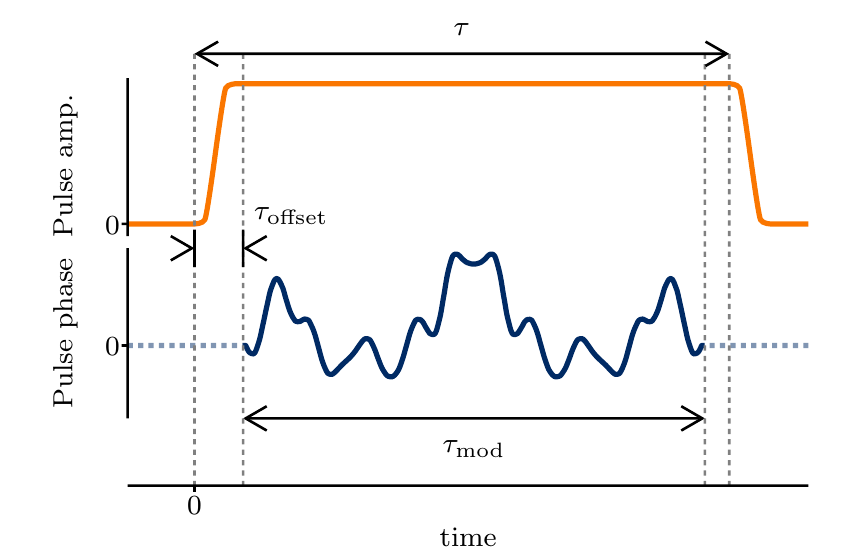}
  \caption{Illustration, not to scale, of the timing parameters optimized to
    realize optimal inversion pulses. The Raman light (top) is turned on at
    $t=0$, and off again at $t=\tau$ by an AOM with a rise-time of $\sim 100~$ns.
    The sample rate of the AWG controlling the phase (bottom) is adjusted to set
    the total duration of the phase waveform $\tau_{\text{mod}}$, and this is
    adjusted together with the trigger delay $\tau_{\text{offset}}$ to achieve
    the optimal peak transfer when $\tau$ is scanned. When the phase waveform
    has finished, the phase is maintained at its final value $\phi_N$.}
  \label{fig:pulse-timings}
\end{figure}

The phase of the RF signal driving the EOM is modulated with a Miteq
SDM0104LC1CDQ I\&Q modulator, the I and Q inputs of which are controlled by the
dual outputs of a Keysight 33612A arbitrary waveform generator (AWG). To realize
a phase sequence $\phi_n, n=1,2,...,N$, the outputs are programmed with the
waveforms $I_n = \sin(\phi_n)$ and $Q_n = \cos(\phi_n)$ and are configured to
hold the final phase value $\phi_N$ until a hardware trigger is received and the
waveforms are played at a sample rate that is adjusted to set the total duration
of the modulation $\tau_{\text{mod}}$ \footnote{In practice, the time required
  for the AWG to respond to a hardware trigger is determined by the sample rate.
  We therefore oversample waveforms, so that each $\phi_n$ comprises many
  ``subsamples'' of equal phase, in order to keep the AWG sample period much
  less than the $\sim 100~$ns rise-time of the AOM that shutters the Raman
  beams.}.

By concurrently triggering the phase modulation AWG and the AOM that shutters
the Raman beams, then measuring the fraction of atoms in the excited state $P_e$
once the AOM shutter has been turned off again after a variable time $\tau$, we
can track the temporal evolution of the atomic state during a phase sequence.
The hardware trigger delay and sample rate of the AWG are adjusted in order to
vary the start time $\tau_{\text{offset}}$ and duration $\tau_{\text{mod}}$ of
the phase waveform respectively (Figure \ref{fig:pulse-timings}) and maximize
the peak excited fraction in these temporal scans.

Having chosen the experimental pulse timings such that the fidelity is maximized at a single value of the Raman detuning $\delta$, set
to the centre of the light-shifted resonance determined from the spectral
profile of a rectangular $\pi$-pulse, a spectral profile can be measured by
measuring $P_e$ upon completion of the pulse at a range of detuning values.

\begin{figure}[t]
  \centering \includegraphics[width=246pt]{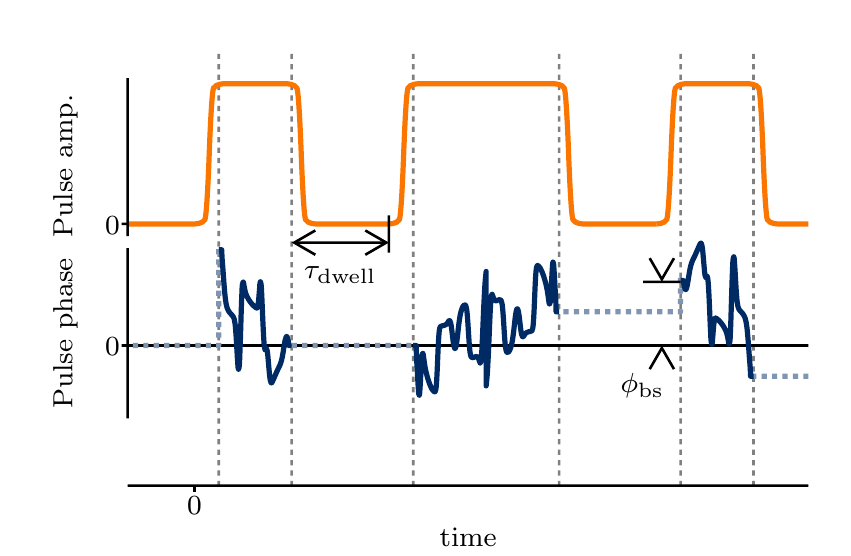}
  \caption{Illustration, not to scale, of an optimized interferometer sequence.
    Three pulses, as depicted in figure \ref{fig:pulse-timings}, are combined,
    with a time $\tau_{\text{dwell}}$ between the light of one pulse turning off
    and the subsequent pulse turning on. Fringes are measured by measuring the
    fraction of atoms in the excited state at the end of the sequence as a
    function of a phase offset $\phi_{\text{bs}}$ applied to the phase sequence
    for the final beamsplitter pulse.}
  \label{fig:interferometer-timings}
\end{figure}

Multiple pulses can be performed sequentially, separated by periods of free
evolution $\tau_{\text{dwell}}$, in order to test optimized interferometer
sequences as illustrated in Figure \ref{fig:interferometer-timings}. The
interferometer contrast is then tested by varying a phase offset
$\phi_{\text{bs}}$ applied to the phase sequence for the final beamsplitter
pulse between 0 to $4 \pi$ and fitting a sinusoidal function to the resulting
fringes in $P_e$.

\section{Results: LMT inversion pulses}\label{sec:exper-real-optim}

We have designed a population inversion pulse using GRAPE that maximizes the
transfer of atoms initially in the state $\ket{\mathrm{g}}$ to the state
$\ket{\mathrm{e}}$ for a cloud with a temperature of 120 $\mu$K and a large
variation in Rabi frequency of $\pm 10\%\ \Omega_{\mathrm{eff}}$.

\begin{figure}[t]
  \centering \includegraphics[width=246pt]{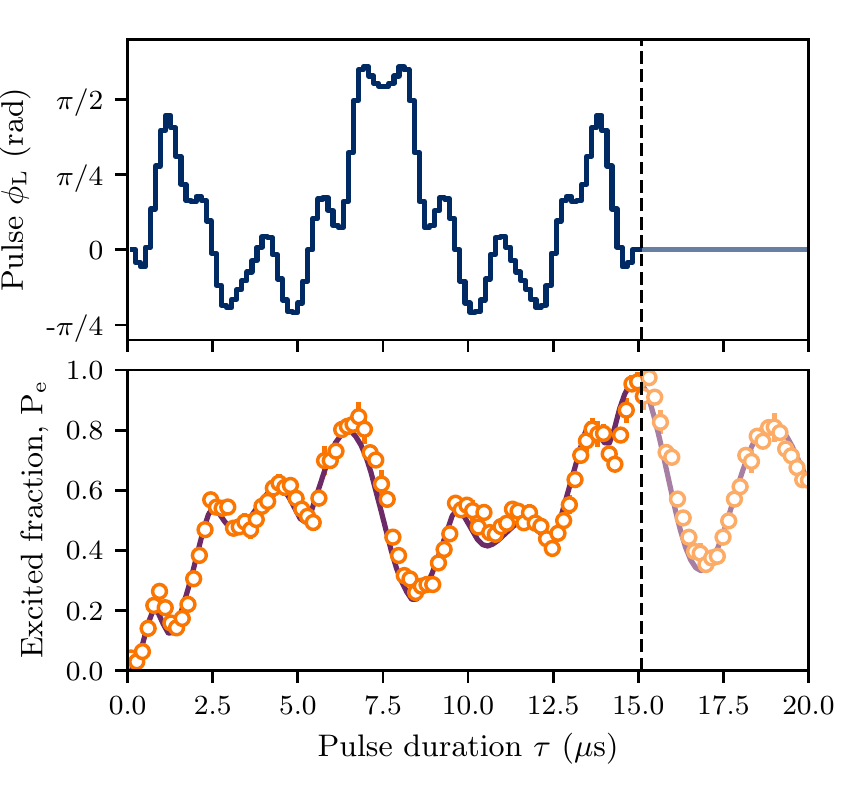}
	\caption{\textit{Above}: Optimal state transfer pulse designed using GRAPE
    to transfer atoms between levels $\ket{\mathrm{g}}$ and $\ket{\mathrm{e}}$.
    \textit{Below}: Measured fraction (circles) of atoms in the excited
    state $|c_{\mathrm{e}}|^2$ after the Raman light is extinguished at
    various times during a pulse. The solid line is a theoretical curve
    produced by the model from \cite{Dunning2014}, in which the two-level
    Hamiltonian is numerically integrated over the range of detunings and
    coupling strengths present in a thermal cloud of $^{85}$Rb atoms, and
    assumes that the light reaches full intensity instantaneously and
    concurrently with the start of the phase modulation. Excellent agreement
    is observed for a simulated temperature of $35~\mu$K.
    \label{fig:temporal-scan}}
\end{figure}

The pulse duration was chosen to be 12 $\mu$s for a Rabi frequency of $310$ kHz,
making it 7.4 times longer than a rectangular $\pi$-pulse, and allowing for a high terminal optimization fidelity. This pulse had 100
timesteps and the algorithm converged to the symmetric waveform shown in Figures
\ref{fig:pulse_profiles} and \ref{fig:temporal-scan} when optimizing the
point-to-point fidelity $\mathcal{F}_{\mathrm{A}}$ with a penalty term added,
proportional to the difference between adjacent pulse steps, to enforce waveform
smoothness \cite{Goodwin2015}. We found that increasing the number of timesteps in this pulse led only to a negligible increase in fidelity.

The temporal profile, after optimizing $\tau_{\text{offset}}$, is shown in Figure \ref{fig:temporal-scan},
showing a peak in the excited fraction $P_e$ at the end of the phase sequence,
represented by the vertical dashed line, after which damped Rabi oscillations are
evident as the phase is kept constant.

We find that optimized pulses demonstrate a considerable resilience to
variations in the trigger delay $\tau_{\text{offset}}$, and can be started as
much as a quarter of a $\pi$-pulse duration after the light is turned on with
little change to the peak transfer.

The highest excited fraction for all pulses is achieved when the light is kept
on for slightly longer than the phase sequence, with $\tau > \tau_{\text{mod}}$
by $\sim 200$~ns. The AOM rise-time is not factored into the optimization
process and this is the only effect of it that we observe experimentally, with
the temporal data being well fitted by a numerical model that assumes the light
reaches full intensity instantaneously at the start of the phase sequence as
shown in Figure \ref{fig:temporal-scan}.
	
\begin{figure*}[t!]
  \subfloat[]{\includegraphics[width=246pt]{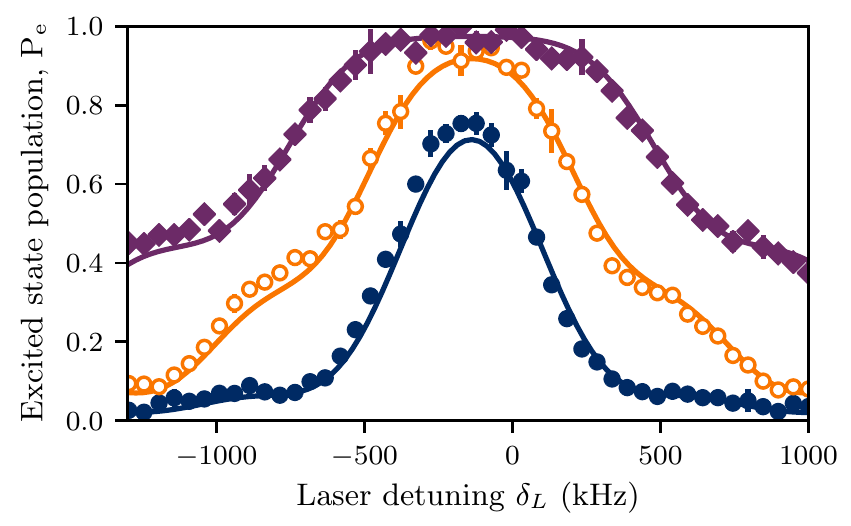}} \hfill
  \subfloat[]{\includegraphics[width=246pt]{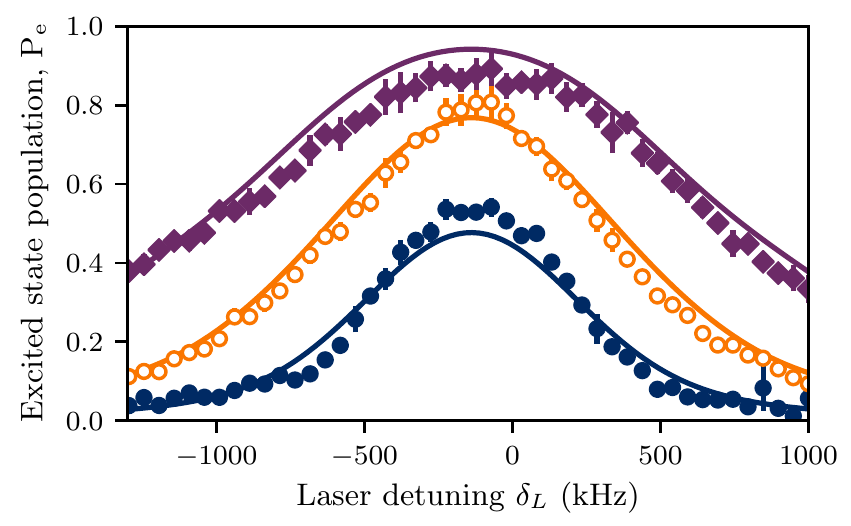}}\\
  \caption{Fraction of atoms transferred to the excited state $P_{\mathrm{e}}$
    as a function of laser detuning in (a) a $35~\mu$K atom cloud and a (b)
    $150~\mu$K cloud. Data are shown for our GRAPE pulse (diamonds), the WALTZ pulse
    (empty circles) and a rectangular $\pi$-pulse (filled circles). The
    effective Rabi frequency was $270$ kHz. Solid lines show theory curves
    produced from the model used by Dunning \textit{et al.} \cite{Dunning2014},
    which assumes a Maxwell-Boltzmann atomic velocity distribution. }
  \label{fig:spectral-scan}
\end{figure*}

To demonstrate the potential of this GRAPE pulse for augmenting LMT
beamsplitters, we compare its performance with the WALTZ composite pulse, the
best composite Raman pulse previously used for LMT interferometry
\cite{Butts2013}, and the standard rectangular $\pi$ pulse over a range of
detunings $\delta$.

In a sub-Doppler cooled cloud of $35~\mu$K, GRAPE and WALTZ pulses achieve close
to $99.8(3)\%$ and $96(2)\%$ transfer respectively about the light-shifted
resonance, while a rectangular $\pi$-pulse achieves just $75(3)\%$. This is
shown in Figure \ref{fig:spectral-scan}, where the broadband nature of the GRAPE
pulse is evident, while the fidelity of the WALTZ pulse drops below $90\%$ when
detuned 100~kHz from resonance, the GRAPE pulse can be detuned by $380$~kHz for
the same fidelity. This broad spectral profile is a signature of a good LMT
augmentation pulse, which must work equally well for atoms that have received a
large number of recoil kicks \cite{Butts2013}.
	
In a cloud with a temperature much closer to the Doppler cooling limit $\sim
150~\mu$K, for which the peak transfer of a rectangular $\pi$-pulse is just $54(2)\%$, the broader spectral width of the WALTZ and GRAPE pulses results in
more efficient state transfer on resonance. This is also shown in Figure
\ref{fig:spectral-scan}, where the GRAPE pulse achieves $89(4)\%$ transfer on
resonance compared with $81(4)\%$ for WALTZ.

\section{Results: Mach-Zehnder interferometer}
	
Using fidelities for optimal beamsplitter and mirror pulses ($ \mathcal{F}_1 $
and $ \mathcal{F}_2$) we optimized all three pulses of the Mach-Zehnder interferometer sequence for
an atomic sample with a temperature of 120 $\mu$K and a coupling strength
variation of $\pm$10\% $\Omega_{\mathrm{eff}}$. The resulting pulse profiles are
shown in Figure \ref{fig:pulse_profiles} (b, c). The phase sequence of the final pulse
was taken to be the inverted and time-reversed profile of the first according to
the design procedure outlined in section \ref{sec:fidelities}. As illustrated in Figure \ref{fig:Contour}, we expect our optimal
Mach-Zehnder sequence of pulses is capable of maintaining a higher contrast than conventional rectangular pulses despite
significant variations in detuning and Rabi-frequency in the atomic cloud. 

\begin{figure}
  \centering \includegraphics[width=246pt]{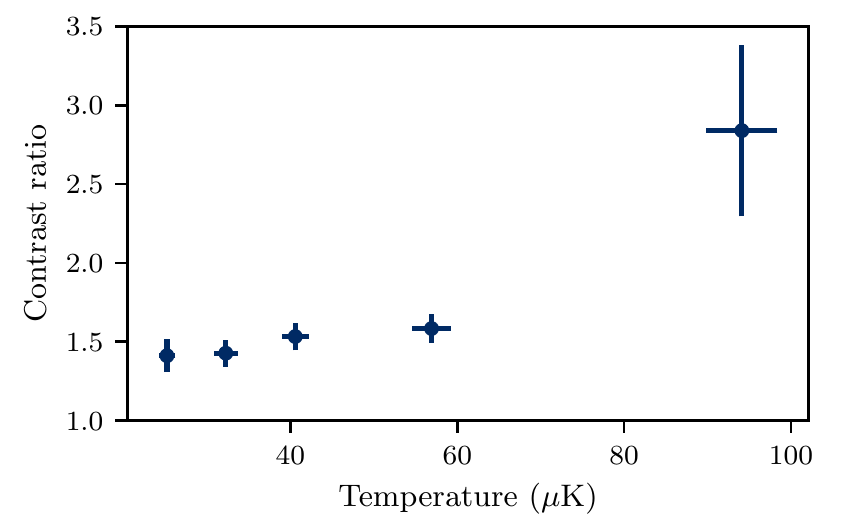}
  \caption{Ratio of contrasts obtained by fitting sinusoidal functions to
    fringes obtained from optimized ``flip-reversed'' GRAPE and rectangular
    interferometer sequences for a range of cloud temperatures. GRAPE
    consistently improves the interferometer contrast with a significant $2.8(6)$ times improvement at the highest cloud temperature of 94(4) $\mu$K
    where temperature and not laser phase noise is the dominant source of
    contrast loss.}
  \label{fig:contrast-ratio}
\end{figure}

\begin{figure}
  \centering \includegraphics[width=246pt]{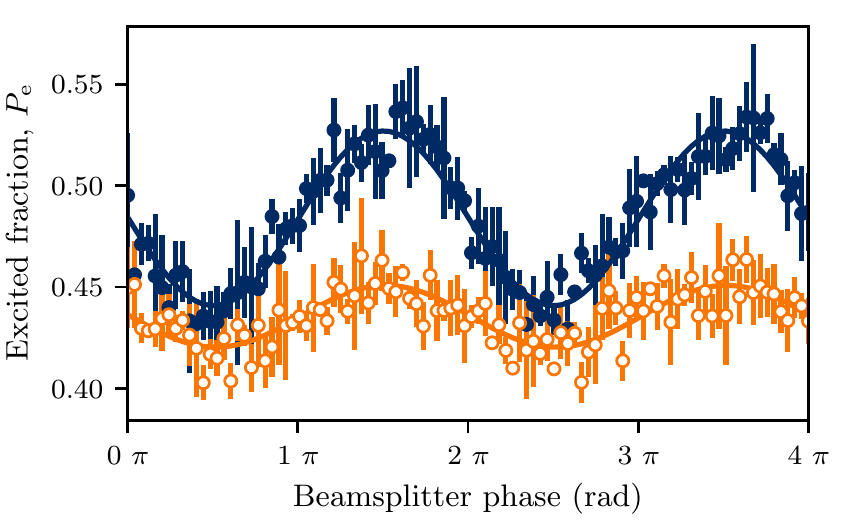}
  \caption{Interferometer fringes obtained at 94(4) $\mu$K for rectangular
    pulses (empty circles) and the optimized GRAPE sequence (filled circles).
    The GRAPE sequence improved the contrast of the fringes by a factor of $2.8(6)$. The average or, `effective' Rabi frequency was approximately $420$
    kHz, and was determined empirically from the optimal duration of a
    rectangular $\pi$ pulse. We attribute the slight deviation of the fringes
    from a sinusoidal form to a small non-linearity in the response of the I\&Q
    modulator.}
  \label{fig:MZ-fringes}
\end{figure}

\begin{figure}
  \centering \includegraphics[width=246pt]{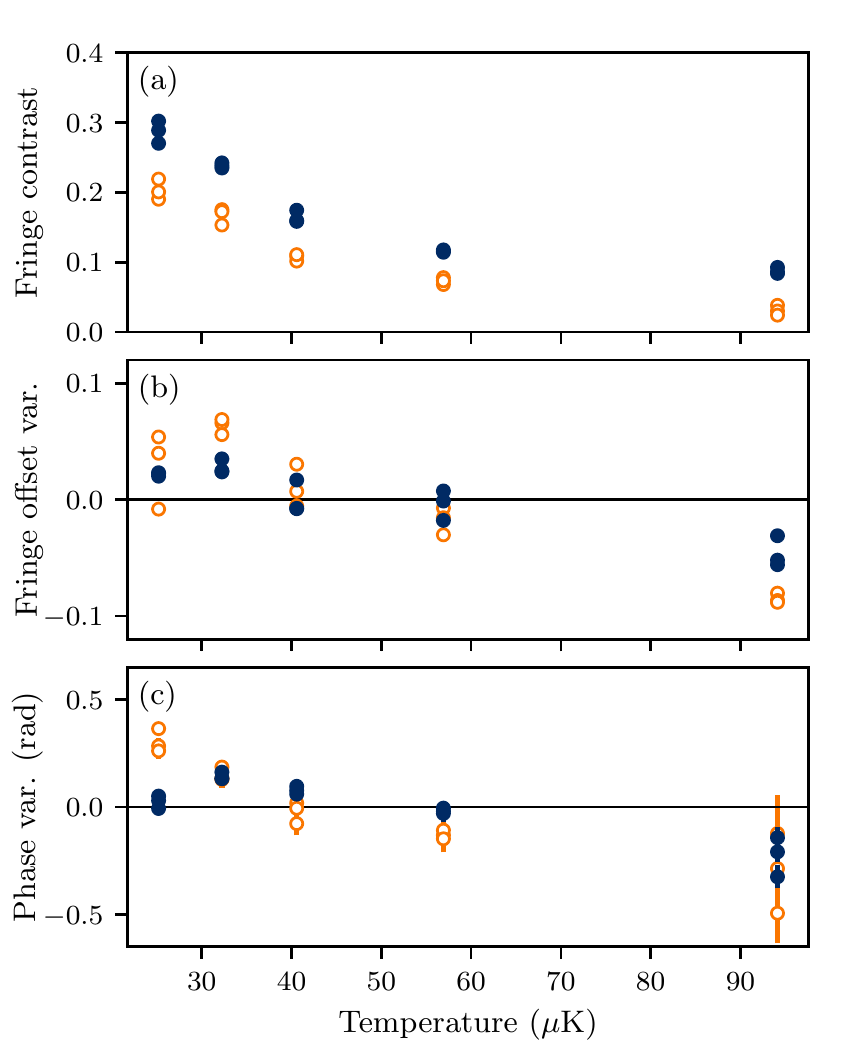}
  \caption{Variation in fitting parameters for sinusoidal functions fitted to
    fringes obtained from GRAPE optimized (filled circles) and rectangular
    (empty circles) interferometer sequences at a range of cloud temperatures.
    (a) shows the fringe contrast $\mathcal{B}$, while (b) and (c) show the
    shifts in the fringe offset $2\mathcal{A}$ and phase from their respective
    mean values. The GRAPE interferometer has consistently higher contrast than
    the rectangular counterpart, and exhibits less variation in both fringe
    offset and phase. Temperature error bars are omitted for clarity, but are
    the same as in Figure \ref{fig:contrast-ratio}.}
  \label{fig:fitting-comparison}
\end{figure}

We have started to test these three-pulse interferometer sequences
in our experimental apparatus, comparing the performance with that obtained 
using a sequence of conventional rectangular $\pi/2$ and $\pi$ pulses. Interferometer fringes are obtained by scanning the phase offset $\phi_{\text{bs}}$ applied to the final pulse in the
sequence. The dwell time between pulses is limited, at present, to $100~\mu$s by
phase noise between the counter-propagating beams that also limits the overall
contrast. The initial results are promising, and the relative improvement in contrast
provided by the GRAPE-optimized sequence as the cloud temperature is varied is shown in
Figure~\ref{fig:contrast-ratio}. GRAPE improved the contrast of the fringes at
all temperatures investigated and, in a hot 94(4) $\mu$K sample where the
contrast loss is dominated by atomic temperature and not the laser phase noise,
nearly a threefold enhancement is observed, the fringes for which are shown in
Figure \ref{fig:MZ-fringes}. This enhancement is also apparent in the uncertainties in the phases of the fitted sinusoids. At 94(4) $\mu$K the average uncertainty in the fitted phase due to the GRAPE sequence was a factor of 2.9 smaller than the uncertainty in the fitted phase from the rectangular pulse sequence.

Some experimental evidence suggests that GRAPE sequences are less susceptible to
drifts in offset and phase of the interferometer fringes. The variation of the
fringe offsets and phases from their respective means for a range of temperatures
for GRAPE and rectangular interferometers are shown in Figure
\ref{fig:fitting-comparison}(b) and \ref{fig:fitting-comparison}(c), and we hope
to explore this aspect more systematically in future work. In particular, there appears to be a systematic shift in the interferometer
phase as the temperature is increased that we do not fully understand, but could
possibly be caused by an offset of the laser detuning from the centre of the
atomic momentum distribution \cite{Gillot2016}. It is notable that the GRAPE
interferometer appears less susceptible to this shift.

Another route of inquiry will be to explain why the increase in contrast is quite
so large only when employing a fully optimised pulse sequence. While the
mirror pulse, with its increased Doppler sensitivity, should be the dominant
source of contrast loss in a Mach-Zehnder interferometer \cite{Saywell2018b},
only a slight enhancement was observed when just this pulse was replaced. To see
significant improvement from a fully optimized sequence, maintaining the overall
anti-symmetry in a ``flip-reversed'' configuration proved necessary. When this constraint was met, the contrast improvement far
exceeded that of replacing just the mirror, or indeed the beamsplitters, in
isolation.

\section{Conclusion}

We have presented the design and experimental implementation of optimal Raman pulses that obtain significant improvements in fidelity compared with conventional pulses. We have used optimal control to design Raman pulses that achieve robust population inversion in the presence of variations in the atom-laser coupling strength and detuning, designed to improve the contrast of LMT interferometers. We have also outlined and demonstrated a design for optimal pulses that improve the contrast of the three-pulse Mach-Zehnder sequence. We expect such optimal pulse sequences to have applications in improving the sensitivity and robustness of atom interferometric sensors
operating in non-ideal environments, relaxing the requirement for low atomic
temperatures and potentially reducing their susceptibility to drifts in the
signal phase and offset.

Specifically, we have presented a `point-to-point' inversion pulse that achieves
99.8(3)\% transfer between hyperfine ground states in a thermal cloud of $^{85}$Rb atoms. Our pulse has a broad spectral profile making it a good
candidate for an augmentation pulse in LMT interferometry where a large velocity
acceptance is required. The best results of Raman LMT interferometers to date
have been achieved with adiabatic augmentation pulses
\cite{Jaffe2018,Kotru2015}, but optimal control pulses such as this have the
potential to realize similarly robust transfer with significant
reductions in pulse duration \cite{Kobzar2004, Goerz2014}.

Furthermore, we have detailed a strategy for optimizing 3-pulse Mach-Zehnder
type interferometer sequences in which optimized beamsplitter and mirror pulses
are combined in a ``flip-reversed'' configuration to maximize the contrast of
interferometer fringes where the phase of atomic superpositions is important.
We have shown up to a threefold contrast improvement in a proof-of-principle
interferometer with hot 94(4) $\mu$K atoms, although our current
investigations have been limited by experimental phase noise.

Future work will extend our experimental study of optimized interferometer
sequences to test their efficacy and robustness when experimental noise is no
longer such a limiting constraint. We are exploring alternative fidelity measures and ways to optimize all pulses concurrently within interferometer sequences. We are also considering both computationally and analytically how to use the framework of
optimal control to engineer robustness to other factors such as laser phase
noise in atom interferometers.

\begin{acknowledgments}
The authors are grateful to David Elcock for his contributions to the stabilisation and characterisation of the optical and RF apparatus. This work was supported by the UK Engineering and Physical Sciences Research Council (Grant Nos. EP/M013294/1 and EP/L015382/1), and by Dstl (Grant Nos. DSTLX-1000091758 and DSTLX-1000097855).
\end{acknowledgments}

Jack Saywell and Max Carey contributed equally to this work.
	
\bibliography{experimental-pulses}

\end{document}